\documentclass[compactaffiliation]{Interspeech}
\usepackage{etoolbox}
\makeatletter
\patchcmd{\affiliation}
  {\renewcommand{\affiliationsep}{\vskip1pt}}
  {\renewcommand{\affiliationsep}{,\space}}{}{}

\makeatother

\interspeechcameraready

\title{EnvSDD: Benchmarking Environmental Sound Deepfake Detection}

\makeatletter

\renewcommand{\authorlist}{%
  Han Yin$^{1,2}$, Yang Xiao$^{3,4}$, Rohan Kumar Das$^{4}$, Jisheng Bai$^{1}$, Haohe Liu$^{5}$ \\
  Wenwu Wang$^{5}$, Mark D Plumbley$^{5}$%
}
\makeatother

\affiliation{}{Northwestern Polytechnical University}{China}
\affiliation{Speech Lab}{Alibaba Group}{China}
\affiliation{}{The University of Melbourne}{Australia}
\affiliation{}{Fortemedia Singapore}{Singapore}
\affiliation{Centre for Vision, Speech and Signal Processing (CVSSP)}{University of Surrey}{UK}

\email{yinhan@mail.nwpu.edu.cn}

\usepackage{comment}
\usepackage{multirow}
\usepackage{graphicx}
\usepackage{url}
\usepackage{hyperref}
\usepackage{booktabs}
\usepackage{amssymb}
\usepackage{bbding}
\usepackage{pifont}
\usepackage{wasysym}
\usepackage{utfsym}
\usepackage{fontawesome}
\usepackage{enumitem}
\usepackage{cite}

\usepackage{colortbl}
\usepackage{array,tabularx}
\usepackage{hyperref}

\definecolor{tablerow1}{RGB}{230,230,230}
\definecolor{tablerow2}{RGB}{245,245,245}

\usepackage[most]{tcolorbox}

\begin{document}

\maketitle

\begin{abstract}
Audio generation systems now create very realistic soundscapes that can enhance media production, but also pose potential risks. Several studies have examined deepfakes in speech or singing voice. However, environmental sounds have different characteristics, which may make methods for detecting speech and singing deepfakes less effective for real-world sounds. In addition, existing datasets for environmental sound deepfake detection are limited in scale and audio types. To address this gap, we introduce EnvSDD, the first large-scale curated dataset designed for this task, consisting of 45.25 hours of real and 316.74 hours of fake audio. The test set includes diverse conditions to evaluate the generalizability, such as unseen generation models and unseen datasets. We also propose an audio deepfake detection system, based on a pre-trained audio foundation model. Results on EnvSDD show that our proposed system outperforms the state-of-the-art systems from speech and singing domains.
    
    
\end{abstract}

\section{Introduction}

Imagine sitting in a college classroom when you suddenly hear a fire alarm start, its sound gradually growing louder. At first, you may think it is a false alarm. Then soon, you hear frantic rush of footsteps and even a siren at a distance. Instinctively, most of you would rise from your seats and begin to move toward the exit. But what if these sound events are not real at all? What if these sounds are created by artificial intelligence (AI)? A similar event was shown in a TikTok video\footnote{\href{https://www.tiktok.com/@andreaspoly/video/7291362107491093803}{https://www.tiktok.com/@andreaspoly/video/7291362107491093803}} that received over 10 million views.
This story shows a new and growing concern in our society: the rise of AI-generated environmental sounds. 
With the growth of open-sourced audio generation models, such as AudioLDM~\cite{audioldm} and AudioLCM~\cite{audiolcm}, it is now becoming easier for people to create very realistic environmental sounds, which previously needed entire teams of professional individuals to produce. 
While these technologies have vast applications in fields such as virtual reality and media production, they also pose serious threats, such as the potential for malicious use in misinformation and the generation of fake audio content that can mislead the public~\cite{audiogen_impact_review}. 
Therefore, these societal concerns show a strong need to develop methods that can detect fake environmental sounds accurately.

The goal of environmental sound deepfake detection (ESDD) is to determine whether a sound clip is sourced from real-life scenarios or has been artificially generated by models. 
Recently, speech and singing voice deepfake detection have received widespread attention, leading to various challenges and benchmarks \cite{asvspoof5,asvspoof2021,svdd2024}. 
However, ESDD poses challenges that are different from those in speech and singing domains. 
For example, speech and singing have specific pitch ranges, but environmental sounds lack steady rhythms and fixed tones. 
In addition, for environmental sounds, many sound events may occur simultaneously, a phenomenon that may not be as apparent in pure speech or singing clips. 
Therefore, we question whether methods from speech and singing deepfake detection can be directly applied to environmental sounds. 


Several studies have focused on fake environmental sound detection.
In SceneFake~\cite{scenefake}, the authors first enhance the real speech involving a scene (e.g., ``Airport''), then add another scene to the enhanced speech (e.g., ``Street''), resulting in a scene-fake audio clip.
Similar to SceneFake, in EnvFake~\cite{envfake}, the authors create a multimodal scene-fake dataset by combining visual and audio information.
These works focus on scene-consistency detection, where all environmental audio clips are sourced from real-life scenarios rather than being generated by AI models.
Ouajdi et al.~\cite{esdd} use real-fake paired audio data from a Foley sound synthesis challenge for ESDD, where fake sounds are generated by 44 submitted audio generation systems.
However, this work only includes monophonic audio, and only conducts in-domain evaluation, without considering the models’ generalization to unseen domains.
In FakeSound~\cite{fakesound}, the authors mask a portion of the audio and use audio inpainting models to generate new audio in the masked segments, creating fake sound clips. 
In this work, most audio clips are polyphonic, with limited consideration of monophonic conditions. 
In addition, they use a relatively small subset of AudioCaps~\cite{audiocaps} for generating fake audio, with just about 5 hours of data.

\begin{figure*}
\centering
\centerline{\includegraphics[width=0.9\textwidth]{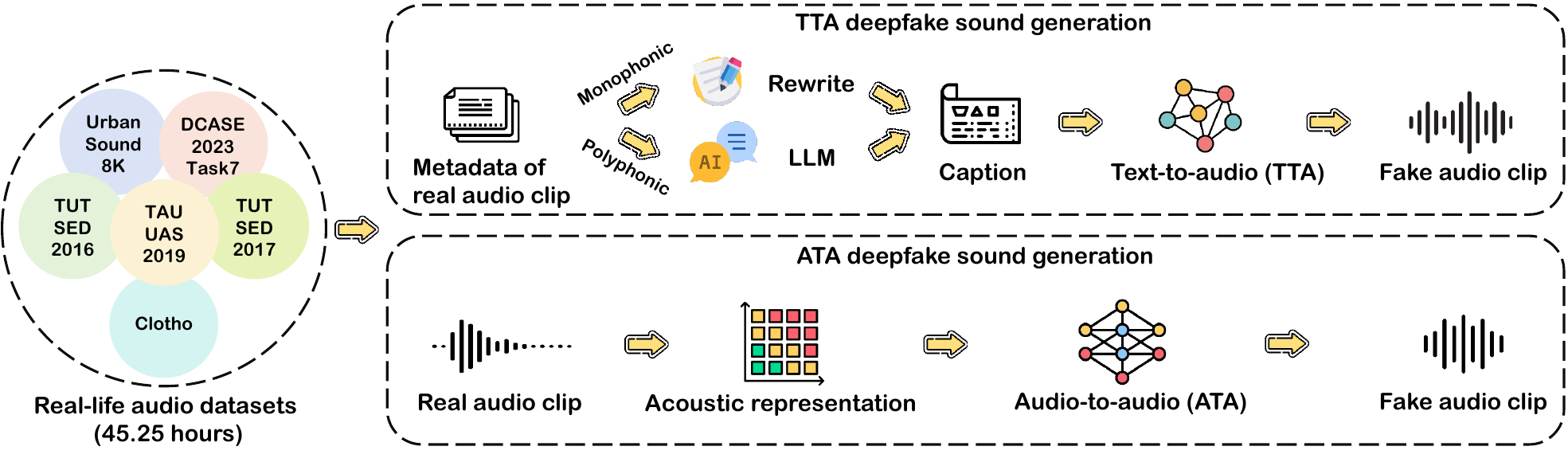}}
\vspace{-2mm}
\caption{Pipeline for creation of the proposed EnvSDD dataset.}
\label{fig:overview}
\vspace{-3mm}
\end{figure*}

In our work, we present EnvSDD, the first large-scale dataset for ESDD, with 45.25 hours of real audio and 316.74 hours of deepfake environmental sound clips. 
We sample real-life sound clips from six datasets, covering both monophonic and polyphonic conditions.
Additionally, we use five state-of-the-art (SOTA) audio generation models to generate the deepfake audio clips. 
Specifically, we consider two types of deepfakes, including the two primary paradigms of audio generation: text-to-audio (TTA) and audio-to-audio (ATA). 
To evaluate the dataset, we first train two SOTA systems from the speech and singing voice deepfake detection domains on EnvSDD.
Then, we propose a new system leveraging a pre-trained audio foundation model (i.e., BEATs~\cite{beats}), to improve the detection performance. The main contributions of this paper are as follows.

\begin{itemize}[left=1.5em]
    \item We present EnvSDD, a large-scale dataset for ESDD.
    \item We propose a new detection system, which outperforms the baselines across all evaluation scenarios. The dataset, models, and codes are all publicly available.\footnote{\href{https://envsdd.github.io}{https://envsdd.github.io}} 
\end{itemize}


\section{EnvSDD Dataset}
\label{sec:dataset}
As shown in Figure~\ref{fig:overview}, the EnvSDD dataset includes two types of deepfakes: TTA and ATA. We first sample real-life audio clips from various publicly available datasets and then generate deepfake clips using the TTA and ATA models. 

\vspace{-1mm}
\subsection{Real Data Collection}
EnvSDD includes both monophonic and polyphonic audio. 
Monophonic audio contains a single event per clip, while polyphonic refers to audio with multiple overlapping sound events. 
These two types are considered to account for the diverse range of real-world scenarios, where both isolated and overlapping sound events occur frequently. For monophonic audio, we use UrbanSound8K~\cite{urbansound8k} and DCASE 2023 Task7 Dev~\cite{dcase2023task7} as the sources of real data. For polyphonic audio, we utilize TAU UAS 2019 Open Dev~\cite{tut_uas}, TUT SED 2016~\cite{tutsed2016}, TUT SED 2017~\cite{tutsed2017_eva,tutsed2017_dev} and Clotho~\cite{clotho}. All audio samples are segmented into 4-second clips, and since most audio generation models are trained with 16 kHz audio, we resample all clips to 16 kHz.

\textit{Monophonic datasets:} \textbf{UrbanSound8K} is a widely used dataset for environmental sound classification, specifically designed for urban sound recognition tasks, which contains 10 different urban sound classes, such as ``siren'' and ``drilling''.
We select audio clips with a duration of 4 seconds and a sampling rate above 16 kHz from it, resulting in 4,523 clips.
\textbf{DCASE 2023 Task7 Dev} refers to the development set of Task 7 from DCASE 2023 challenge, which contains real-life audio clips from various public resources with 7 event classes. 

\textit{Polyphonic datasets:} \textbf{TAU UAS 2019 Open Dev} is designed for acoustic scene classification, consisting of audio clips recorded in 14 different real-life acoustic scenes, such as ``airport'' and ``office''.
\textbf{TUT SED 2016} and \textbf{TUT SED 2017} are two environmental audio datasets for sound event detection, containing real-life recordings from various indoor and outdoor scenarios.
\textbf{Clotho} is a dataset designed for audio captioning tasks, which contains audio clips from FreesSound~\cite{freesound}, and human-written captions describing the audio content.

Since DCASE 2023 Task 7 Dev and Clotho are used only for evaluation, we select a small subset from these datasets, covering various real-life scenarios. 
Additionally, the ``DogBark'' and ``GunShot'' events in DCASE 2023 Task 7 Dev overlap with UrbanSound8K, and are therefore excluded.

\subsection{Deepfake Data Generation}
In the proposed EnvSDD dataset, pre-trained TTA and ATA models are applied to generate deepfake audio clips. As shown in Table \ref{tab:attack_model}, we use five models for TTA and two models for ATA, with an inference step of 100.

\begin{table}
  \caption{Overview of audio generation models used in the proposed EnvSDD dataset.}
  \vspace{-2mm}
  \label{tab:attack_model}
  \centering
    \renewcommand\arraystretch{1.2}{
\setlength{\tabcolsep}{6mm}{
  \begin{tabular}{c|c}
    \toprule
    \textbf{Deepfake Type} & \textbf{Generation Model (G\#)} \\
    \midrule
    \multirow{5}{*}{Text-to-Audio} & AudioLDM (G1)\\
    & AudioLDM 2 (G2)\\
    & AudioGen (G3)\\
    & TangoFlux (G4)\\
    & AudioLCM (G5)\\
    \midrule
    \multirow{2}{*}{Audio-to-Audio} & AudioLDM (G1)\\
    & AudioLDM 2 (G2)\\
    \bottomrule
  \end{tabular}
  }}
  \vspace{-5mm}
\end{table}

\begin{table*}[t!]
  \caption{Statistics of the proposed EnvSDD dataset.}
  \vspace{-2mm}
  \label{tab:audio_clips}
  \centering
      \renewcommand\arraystretch{1}{
    \setlength{\tabcolsep}{7mm}{
  \begin{tabular}{c|c|ccc}
    \toprule
    \textbf{Audio Type}      &\textbf{Source Dataset (D\#)} & \textbf{\#Real} & \textbf{\#Fake (TTA)} & \textbf{\#Fake (ATA)} \\
    \midrule
    \multirow{2}{*}{Monophonic} & UrbanSound8K (D1) & 4,523 & 22,615 & 9,046 \\
    & DCASE 2023 Task7 Dev (D2)& 500 & 2,500 & 1,000 \\
    \midrule
    \multirow{4}{*}{Polyphonic} & TAU UAS 2019 Open Dev (D3)& 31,700 & 158,500 & 63,400 \\
    & TUT SED 2016 (D4) & 1,695 & 8,475 & 3,390 \\
    & TUT SED 2017 (D5) & 1,806 & 9,030 & 3,612 \\
    & Clotho (D6) & 500 & 2,500 & 1,000 \\
    \bottomrule
  \end{tabular}
  }
  \vspace{-5mm}
  }
  
\end{table*}

\textit{TTA Deepfake:} We used 5 models for generating TTA deepfake clips, namely, AudioGen~\cite{audiogen}, AudioLDM~\cite{audioldm}, AudioLDM 2~\cite{audioldm2}, TangoFlux~\cite{tangoflux} and AudioLCM~\cite{audiolcm}. An important challenge is how to generate accurate and meaningful captions for audio clips, which are applied as inputs to the TTA models.

As shown in Figure \ref{fig:overview}, for monophonic audio, we directly rewrite the metadata as the audio caption. ``Rewrite'' refers to removing meaningless characters from the metadata, capitalizing the first character, and adding a period in the end (e.g., ``gun\_shot'' is rewritten to ``Gun shot.''). For polyphonic audio, a large language model (LLM) is applied to generate captions based on the metadata. Specifically, we use Mistral 7B~\cite{mistral} as the LLM, with prompts configured as follows. If the metadata contains both scene and event labels, the prompt is ``Prompt A'' here as shown in the box:

\vspace{-1mm}
\begin{tcolorbox}[halign title=flush center, title={Prompt A},title filled, 
                  colback=white,
                  colframe=gray!80,
                  boxrule=0.8pt,
       tabularx*={\renewcommand*{\arraystretch}{1.3}}%
                 {>{\raggedright\arraybackslash\hsize=\hsize}X%
              }
              ]
              
    \arrayrulecolor{white}
    
      This clip is recorded in \{scene label\}, where the following events or sounds are happening: \{event labels\}. \\ A caption is a descriptive sentence, which vividly depicts the acoustic content of the audio clip. Please provide one sentence for the caption to directly describe the sound.
\end{tcolorbox}


\noindent If the metadata contains only the scene label, the prompt is: 

\vspace{-1mm}

\begin{tcolorbox}[halign title=flush center, title={Prompt B},title filled, 
                  colback=white,
                  colframe=gray!80,
                  boxrule=0.8pt,
       tabularx*={\renewcommand*{\arraystretch}{1.3}}%
                 {>
                 {\raggedright\arraybackslash\hsize=\hsize}X%
              }
              ]
              
    \arrayrulecolor{white}
    
    This clip is an audio clip recorded in \{scene label\}. \\ A caption is a descriptive sentence, which vividly depicts the acoustic content of the audio clip. Please provide one sentence for the caption to directly describe the sound that might occur in the scene.
\end{tcolorbox}

 
\noindent By prompting Mistral 7B, we generate a caption for each audio clip. It is noted that Clotho already provides audio captions, so we do not generate captions for it.

\textit{ATA Deepfake:} 
AudioLDM and AudioLDM 2 are used for ATA, where a latent diffusion model conditions on acoustic representations to generate audio events resembling those in the input. AudioLDM relies on CLAP~\cite{clap} features, while AudioLDM 2 leverages AudioMAE~\cite{audiomae}, which captures finer acoustic details. 
Therefore, ATA with AudioLDM 2 produces audio that more closely matches the original, while AudioLDM generates semantically similar but acoustically distinct outputs.

\subsection{Dataset Splits}

For clarity, as shown in Table \ref{tab:attack_model} and Table \ref{tab:audio_clips}, we label the source datasets as D1-D6 and the generation models as G1-G5. Table \ref{tab:split} presents statistics for different dataset splits. To comprehensively evaluate detection performance across various scenarios, we define four distinct test conditions. 

Test 01 is for in-domain evaluation, where the source datasets and audio generation models are seen during training. 
Test 02-04 are designed for out-of-domain evaluation. 
Specifically, Test 02 evaluates generalizability by using generation models that are different from those during training, assessing the detection model’s ability to handle unseen audio generation models. 
Test 03 focuses on generalizability to unseen datasets by ensuring the source datasets are significantly different from those used in training and validation. Test 04 presents the most challenging scenario, as both the source datasets and generation models differ from those during training.

\begin{table}[t!]
    \centering
    \caption{Statistics of train, validation, and test sets in EnvSDD.}
    \vspace{-3mm}
    \label{tab:split}
    \renewcommand\arraystretch{1.0}
    \setlength{\tabcolsep}{2pt}
    \begin{tabular}{c|c|c|c c}
        \toprule
        \textbf{Split} & \textbf{Source Datasets} & \textbf{Generation Model} & \textbf{\#Real} & \textbf{\#Fake} \\
        \midrule
        \rowcolor{gray!20} \multicolumn{5}{c}{\textbf{TTA Deepfake}} \\
        \midrule
        Train & D1, D3, D4, D5 & G1, G2, G3 & 27,811 & 83,433 \\
        Valid & D1, D3, D4, D5 & G1, G2, G3 & 7,942 & 23,826 \\
        \cmidrule{1-5}
        Test 01 &  D1, D3, D4, D5 & G1, G2, G3 & 3,971 & 11,913 \\
        Test 02 & D1, D3, D4, D5 & G4, G5 & 3,971 & 7,942 \\
        Test 03 & D2, D6 & G1, G2, G3 & 1,000 & 3,000 \\
        Test 04 & D2, D6 & G4, G5 & 1,000 & 2,000 \\
        \midrule
        \rowcolor{gray!20} \multicolumn{5}{c}{\textbf{ATA Deepfake}} \\
        \midrule
        Train & D1, D3, D4, D5 & G1 & 27,811 & 27,811 \\
        Valid & D1, D3, D4, D5 & G1 & 7,942 & 7,942 \\
        \cmidrule{1-5}
        Test 01 &  D1, D3, D4, D5 & G1  & 3,971 & 3,971 \\
        Test 02 & D1, D3, D4, D5 & G2 & 3,971 & 3,971 \\
        Test 03 & D2, D6 & G1 & 1,000 & 1,000 \\
        Test 04 & D2, D6 & G2 & 1,000 & 1,000 \\
        \bottomrule
    \end{tabular}
    \vspace{-5mm}
\end{table}

\section{Model Architectures}



\subsection{Baselines: AASIST and W2V2+AASIST}

\noindent Recent studies in speech deepfake detection show that key clues can appear in both the spectral and temporal domains~\cite{tak2021end}. AASIST~\cite{aasist} is an end-to-end system that uses a novel heterogeneous stacking graph attention layer~\cite{gat} to learn these features.
This model has been applied as the baseline in various speech and singing voice deepfake detection challenges~\cite{asvspoof2021,asvspoof5,svdd2024}.
Building on AASIST, W2V2+AASIST~\cite{xlsaasist} uses a pre-trained speech foundation model, wav2vec 2.0 XLS-R~\cite{xls}, as the front-end, to extract representations from the waveform,
achieving SOTA results on ASVspoof2021~\cite{asvspoof2021} and SingFake datasets~\cite{svdd2024}.
The wav2vec 2.0 XLS-R (W2V2) model was pre-trained on 128 languages and approximately 436K hours of unlabeled speech data with self-supervised learning.
We apply these two systems as the baselines, to explore whether methods from speech and singing domains can be applied to environmental sounds.

\subsection{Proposed BEATs+AASIST System for ESDD}

\noindent While W2V2+AASIST works well for speech deepfakes and shows the benefits of a pre-trained speech foundation model, we hypothesize that such a model may not generalize well for diverse sound scape in environmental sounds.
Therefore, we propose a novel system that integrates an audio foundation model with AASIST. We select the audio foundation model due to its pre-training on large-scale, diverse sound datasets, enabling it to capture robust and rich acoustic features for environmental audio.
In our work, we use BEATs~\cite{beats}, a pre-trained model that learns deep audio representations from AudioSet-2M~\cite{audioset} with self-supervised learning. 
BEATs begins with a random projection as an acoustic tokenizer for mask and label prediction, then refines the tokenizer with its learned knowledge in repeated cycles. As in W2V2+AASIST, the output of BEATs is fed to a RawNet2-based residual encoder~\cite{rawnet2} that learns higher-level feature representations. Then a self-attention aggregation layer is applied to extract the acoustic representation, which is fed to the AASIST model to obtain a two-class prediction.

By replacing the W2V2 front-end with BEATs, our proposed system is expected to capture the complex nature of environmental sounds more precisely. 
This integration of BEATs with AASIST is anticipated to enhance performance in detecting environmental audio deepfakes. We evaluate all the three systems on our proposed EnvSDD dataset.

\section{Experiments and Results}
\label{sec4}
\subsection{Experimental Setup}
During training and fine-tuning, we use a batch size of 32 and the Adam optimizer with a weight decay of 0.0001. When training AASIST from scratch, the initial learning rate is 0.001. For fine-tuning W2V2+AASIST and BEATs+AASIST, we reduce the learning rate to 0.00001 to avoid overfitting. The maximum number of epochs is 50. Training stops early if the validation loss does not decrease for 5 consecutive epochs.

We consider equal error rate (EER)~\cite{eer} as the evaluation metric, which is also common in speech and singing deepfake detection. Each system produces a score for a given audio clip, indicating the confidence that the clip is real. The EER is found by setting a threshold where the false acceptance rate equals the false rejection rate. Lower EERs indicate better detection performance of a system.


\begin{table}[t!]
  \caption{Performance in EER (\%) of baseline and proposed systems on EnvSDD dataset.
  ``Seen SD'' means ``Seen Source Datasets'' and ``Seen GM'' means ``Seen Generation Model''.} 
  \vspace{-2mm}
  \label{tab:result1}
  \centering
  \renewcommand\arraystretch{0.95}{
\setlength{\tabcolsep}{0.8mm}{
  \begin{tabular}{c|c|cc|cc}
    \toprule
    \multirow{2}{*}{\textbf{System}} & \multirow{2}{*}{\textbf{Test Set}} & \multicolumn{2}{c|}{\textbf{Test Condition}} & \multicolumn{2}{c}{\textbf{Fake Type}} \\
    & & Seen SD & Seen GM & TTA & ATA\\
    \midrule
    \multirow{5}{*}{AASIST} & Test 01 & \ding{51} & \ding{51} & 0.66 & 0.28\\ 
    & Test 02 & \ding{51} & \ding{55} & 3.70 & 0.68\\
    & Test 03 & \ding{55} & \ding{51} & 6.80 & 3.00\\
    & Test 04 & \ding{55} & \ding{55} & 17.50 & 4.40\\
    & \cellcolor{gray!20} Average & \cellcolor{gray!20} - & \cellcolor{gray!20} - & \cellcolor{gray!20} 7.17 &\cellcolor{gray!20} 2.09 \\
    \midrule
    \multirow{5}{*}{W2V2+AASIST} & Test 01 & \ding{51} & \ding{51}  & 0.26 & 0.38\\
    & Test 02 & \ding{51} & \ding{55}  & 13.04 & 26.59\\
    & Test 03 & \ding{55} & \ding{51}  & 10.60 & 13.30\\
    & Test 04 & \ding{55} & \ding{55}  & 45.80 & 52.40\\
    & \cellcolor{gray!20}Average & \cellcolor{gray!20} - & \cellcolor{gray!20} - & \cellcolor{gray!20}17.43 & \cellcolor{gray!20}23.17 \\
    \midrule
    \multirow{5}{*}{BEATs+AASIST} & Test 01 & \ding{51} & \ding{51} & 0.08 & 0.03\\
    & Test 02 & \ding{51} & \ding{55}  & 1.26& 0.08\\
    & Test 03 & \ding{55} & \ding{51}  & 4.70 & 2.20\\
    & Test 04 & \ding{55} & \ding{55}  & 17.20 & 3.00\\
    & \cellcolor{gray!20}Average & \cellcolor{gray!20} - & \cellcolor{gray!20} - & \cellcolor{gray!20}\textbf{5.81} & \cellcolor{gray!20}\textbf{1.33}\\
    \bottomrule
  \end{tabular}
  \vspace{-5mm}
  }}
\end{table}

\subsection{Performance of Baselines and Proposed System}

We first focus on the performance of the two baselines AASIST and W2V2+AASIST in Table~\ref{tab:result1} under all test conditions. Although both models perform well under the seen condition of Test 01, their performance drops significantly under the unseen test conditions of Tests 02-04. It should also be noted that the introduction of W2V2 in AASIST generally improves the performance of speech deepfake detection as reported in~\cite{xlsaasist}, but the same trend is not observed for deepfake sound detection from the results observed in Table~\ref{tab:result1}. This may be due to the fact that W2V2 was pre-trained on pure speech data, and speech exhibits significant differences in acoustic features compared to environmental sounds. 

We now compare the performance of our proposed BEATs+AASIST system with the two baselines as reported in Table~\ref{tab:result1}. It is evident that the introduction of BEATs significantly benefits the AASIST system, which was pre-trained on environmental sounds.
This suggests that it is able to capture acoustic nuances more effectively than speech-based pre-trained models. 
BEATs+AASIST outperforms both baseline models across all test conditions, validating our hypothesis of leveraging an audio foundation model.



Let us consider having a closer look at the performances of all the systems under each test condition. The low EERs (below 1\%) for all systems in Test 01 indicates that in-domain sound deepfake detection is a relatively simple task. In Test 02, the trained baselines have to handle fake data generated by unseen TTA and ATA models. The results show a significant drop in performance on Test 02 compared to Test 01, demonstrating the detection model's limited ability to generalize across different audio generation models. Similarly, in Test 03, the source datasets are out-of-domain for the trained detection models, leading to degraded performance compared to Test 01. Furthermore, we observe that AASIST and BEATs+AASIST systems are more affected by unseen source datasets than unseen audio generation models, whereas W2V2+AASIST shows an opposite trend indicating more dependence on generation models. Test 04 is the most challenging condition as both the audio generation models and the source datasets are out-of-domain, resulting significant drop in performance for all the systems. However, the proposed BEATs+AASIST system still performs substantially better than the two baselines.




\subsection{Performance for Monophonic and Polyphonic Audio}
\begin{figure}[t!]
\centering
\centerline{\includegraphics[width=0.41\textwidth]{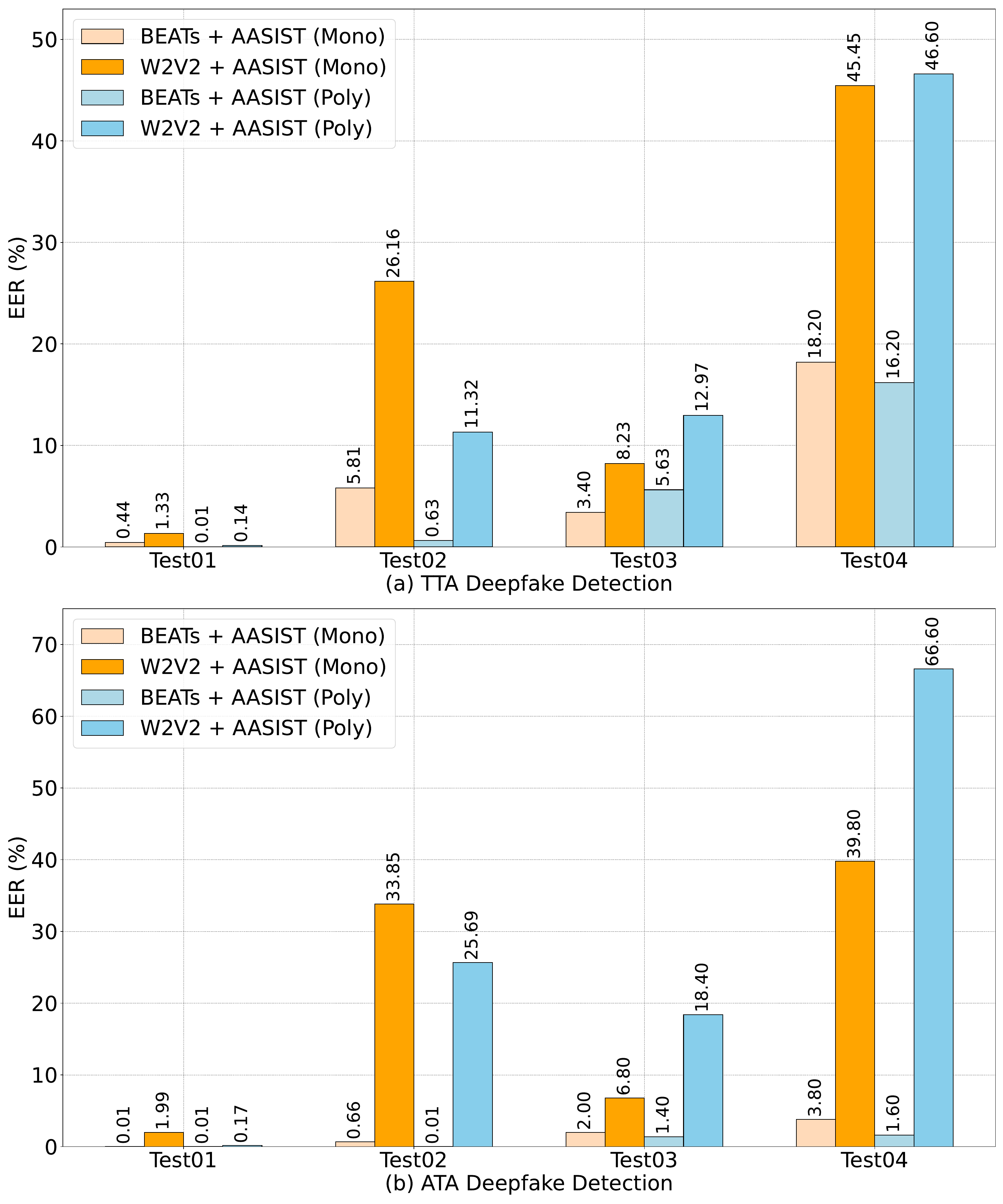}}
\vspace{-2mm}
\caption{Performance in EER (\%) of W2V2+AASIST and BEATs+AASIST on monophonic and polyphonic data.}
\vspace{-6mm}
\label{fig:exp2}
\end{figure}

In this work, the EnvSDD dataset contains two types of audio, monophonic and polyphonic, where the former has only a single sound event, while the latter exhibits multiple overlapping events. Figure~\ref{fig:exp2} shows the detection performance of W2V2+AASIST and BEATs+AASIST on monophonic and polyphonic subsets. For BEATs+AASIST, the EERs on monophonic audio are generally higher than those on polyphonic audio, indicating that it is more challenging to detect monophonic deepfake audio. 
However, the EER on monophonic audio is lower than that on polyphonic audio for TTA deepfake detection in Test 03, where the source datasets are unseen during training.
For W2V2+AASIST, the model performs better on polyphonic audio in Test 01 and Test 02, whereas the opposite trend is observed in Test 03 and Test 04, with better performance on monophonic audio. 
Overall, the proposed BEATs+AASIST outperforms W2V2+AASIST in all cases, further demonstrating the effectiveness of the pre-trained audio foundation model.

\section{Conclusions}
In this work, we have proposed EnvSDD, the first large-scale dataset with real and deepfake environmental audio clips for ESDD. We use two SOTA systems from the speech and singing voice deepfake detection domains as baselines and proposed a new system based on a pre-trained audio foundation model, BEATs. The results on EnvSDD under various test conditions demonstrate that the proposed system outperforms the two baselines. 
However, its generalization to unseen domains remains limited, which should be further explored in future work.

\section{Acknowledgements}
This work was supported by the Engineering and Physical Sciences Research Council [grant numbers EP/T019751/1, EP/Y028805/1]; a PhD scholarship from the Centre for Vision, Speech and Signal Processing (CVSSP), Faculty of Engineering and Physical Science (FEPS), University of Surrey; British Broadcasting Corporation Research and Development (BBC R\&D); and an Adobe Research Gift. For the purpose of open access, the authors have applied a creative commons attribution (CC BY) licence to any author accepted manuscript version arising.

\bibliographystyle{IEEEtran}
\bibliography{mybib}

\end{document}